TITLE PAGE

Title: Machine Learning as a Catalyst for Value-Based Health Care

Authors:

Matthew G. Crowson MD, MPA FRCSC[1,2] – *Corresponding Author*

Timothy C. Y. Chan PhD[2]

1 Department of Otolaryngology-Head & Neck Surgery, Sunnybrook Health Sciences Center, Toronto, Ontario.

2 Department of Mechanical & Industrial Engineering, University of Toronto, Toronto, Ontario

Manuscript Submission:

(1) Each of the authors indicated above have contributed to, read and approved this manuscript.

(2) **FINANCIAL DISCLOSURE**: no authors have disclosures related to this manuscript.

(3) **CONFLICT DISCLOSURE**: no authors have conflicts related to this manuscript.

(4) In consideration of the journal reviewing and editing my submission, the authors undersigned transfers, assigns and otherwise conveys all copyright ownership in the event that such work is published.

Corresponding Author:

Matthew G. Crowson MD MPA FRCSC
Sunnybrook Health Sciences Center
University of Toronto
matt.crowson@mail.utoronto.ca
Toronto, Ontario, Canada

Recent estimates of waste in health care spending in the United States have reached a threshold of one trillion dollars (Shrank et al. 2019), equivalent to the gross domestic product of the 17th largest economy in the world. The major contributors to wasteful spending span failures of care delivery, over-treatment and low-quality care, fraud, administrative complexity and redundancies, inadequate care coordination, and widely variable resource pricing (Shrank et al. 2019).

Determining the factors that drive wasteful spending is a complex problem since waste can arise anywhere from daily individual decisions made by clinicians and their allies to decisions made by health authorities and legislators at the regional or national systems-level. Considering only individual clinical encounters between clinicians and patients, high quality care demands seamless integration of rapidly advancing clinical knowledge, compassionate care, and electronic data all within a fifteen-minute visit. Efficiently securing a diagnosis, work-up, and treatment plan consistent with today's gold standard of care in this continuously evolving and demanding environment increases the probability of error. An error in this pathway such as a missed diagnosis, an inappropriate imaging study, or an ineffective treatment adds to the waste tally. Ultimately, the complexity of modern medicine leads to error resulting in waste, which leads to poor health care value.

The complexity of modern health care delivery and the dire consequences of persistent wasteful care have led to the rising popularity of value-based care models, which are centered on the principle that optimal health care value is achieved when patient outcomes are maximized per unit cost to deliver those outcomes (Porter and Teisberg 2006). If we can improve clinical outcomes without increasing costs, or preferably at lower cost, then we progress towards the goal of higher value care. Several value-inspired delivery models have emerged including regional patient-centered care homes, system-wide accountable care organizations, and bundled payments for diseases or intervention packages. Despite their promise, these solutions have yet to address the complexity of individual decision making in a large lake of data. The probability of misdiagnosis and error rises with increased complexity and uncertainty, which ultimately reinforces barriers to high value care. Error is a key unifying thread among the major contributors to wasteful spending. Reduced error rates in clinical decision making would lead to a reduction in inappropriate care delivery, fewer delays in receiving appropriate care, and improved care coordination.

Despite world-class medical education and significant resources devoted to cutting-edge care, over 80,000 patients die in the United States annually due to diagnostic error (Newman-Toker and Pronovost 2009). These errors may arise, in part, from a failure to integrate the best-available data for every individual clinical decision. Our knowledge of human physiology and pathology rapidly advances year over year, as do our methods for collecting and analyzing clinical data. We have attempted to disseminate new knowledge and tools by introducing evidence-based guidelines, checklists, and simplified rules-based algorithms. Clinicians are called upon to make predictions from all of these inputs at the point of care, but these predictions are subject to error because of our limited capacity to integrate all the data. Our analytic and cognitive plateau needs a new approach that is capable of generating accurate predictions and actionable insights from overwhelming amounts of complex data. New approaches that improve prediction accuracy in the face of complex and dynamic datasets should decrease error and, ultimately, health care waste.

Machine learning has tremendous potential to help reduce error, curtail waste and ultimately improve value through generating accurate clinical predictions. Recent successes have provided international validation of machine learning algorithms applied to diagnostic tasks in medical imaging with classification accuracy performance that rivals or exceeds human performance (McKinney et al. 2020). Successful diagnostic algorithms have also shown promise in the fields of dermatology and pathology. Automatic patient safety event flagging from clinical documentation through the flagging of important index events (e.g., falls, missed medications, or trends in persistent caregiver concerns) could prevent costly sequalae of compromised patient safety in both inpatient and outpatient settings. Local, real-time social media and institutional data could be combined to forecast institutional resource needs to combat seasonal spikes in infectious disease outbreaks or other public health events. Machine learning can also be used to predict optimal patient treatment regimens based on relevant demographics, clinical presentation factors and genomic data. Tailored treatment algorithms would reduce unnecessary or inappropriate care, and maximize the probability for favorable patient outcomes. Additional opportunities are provided in Table 1, aligned with major categories of waste.

| Waste Domain ($, billions) | Example | Algorithm Opportunities |
|---|---|---|
| i. **Administrative Complexity** *(265.5)* | ▪ Redundant or inappropriate billing/coding practices<br><br>▪ Excess or redundant administrative personnel | ▪ Automatic coding based on clinical narratives and reports with natural language processing (e.g., patient complexity coding)<br><br>▪ Health resource forecasting and optimization based on historical trends integrated with live social media sentiments (e.g., real-time prediction of seasonal infectious disease outbreaks) |
| ii. **Pricing Failures** *(230.7-240.5)* | ▪ Costly medication research & development<br><br>▪ Regional variability in resource pricing | ▪ Pattern recognition for accelerated drug discovery<br><br>▪ Automated supplier price anomaly/outlier detection |
| iii. **Care Delivery Failures** *(102-165.7)* | ▪ Preventable hospital-associated adverse events<br><br>▪ Failure to capture preventative care opportunities | ▪ Automatic patient safety event flagging from clinical documentation (e.g., flagging of inpatient falls, missed medications, or nursing concerns)<br><br>▪ Automatic chronic disease risk factor identification (e.g., individualized time-series analysis of biomarkers and compliance with therapies and/or guidelines) |
| iv. **Overtreatment, inappropriate care** *(75.7-101.2)* | ▪ Misdiagnosis<br><br>▪ Improper testing or therapeutic indications | ▪ Automated radiographic and pathological sample diagnosis (e.g., error reduction in missed or incorrect specimen diagnosis)<br><br>▪ Individualized treatment recommendation engine using best available data (e.g., automatic recommendation of optimal testing strategies) |
| v. **Fraud** *(58.5-83.9)* | ▪ Misrepresentation of illness or disability | ▪ Billing and utilization pattern anomaly detection (e.g., risk factor identification of anomalous use of health care resources) |
| vi. **Care Coordination Failure** *(27.2-78.2)* | ▪ Preventable admissions for post-care or post-procedure complications<br><br>▪ Failure of coordination of home medical equipment and resource needs | ▪ Readmission risk prediction (e.g., predictive analysis based on the confluence of clinical narrative, biomarkers, and imaging data)<br><br>▪ Automated home medical equipment identification and sourcing (e.g., prediction for utilization of home mobility assistance needs) |

**Table 1**. Opportunities for machine learning applications to address challenges in wasteful health care spending. Contribution to wasteful spending estimates adapted from Shrank et al., 2019.

Despite the promise of machine learning in medicine, there are significant limitations that have prevented widespread adoption. One of the major roadblocks in the current application of machine learning algorithms is the availability of sufficient high-quality data. Collecting, curating and labeling data is time consuming and expensive. Another key limitation is the interpretability of machine learning algorithm output. Clinicians may be wary to trust and implement a predictive model without knowing the underlying mechanism for how the algorithm arrived at its prediction. More recently, key ethical issues including health information privacy, industry relationships, bias and accountability for machine error have also been raised as important checkpoints that must be addressed (Char et al. 2018). Fortunately, these limitations are active areas of research.

The complexity of modern medicine, error and subsequent wasteful care is one of the key drivers of runaway health care spending in today's health care ecosystem. Value-based health care provides a framework for developing innovations with an aim to achieve higher quality at lower cost. Machine learning is poised to help reduce uncertainty in clinical decision making to serve the value-based health care mandate. Predictive models may also reveal modifiable health system factors to address care delivery shortcomings. While there are many challenges to operationalizing machine learning algorithms globally, the question remains whether we can afford not to.